\documentclass[aps,prl,reprint]{revtex4-1}

\usepackage{graphicx}% Include figure files
\usepackage{amsmath}
\usepackage{textcomp} %texterweiterungen
\usepackage{xcolor}
\usepackage{ulem}

\begin{document}

\title{Temporal behavior of the inverse spin Hall voltage in a magnetic insulator-nonmagnetic metal structure}

\author{M.~B.~Jungfleisch}
\email{jungfleisch@physik.uni-kl.de}
\affiliation{Fachbereich Physik and Forschungszentrum OPTIMAS, Technische Universit\"at Kaiserslautern, 67663
Kaiserslautern, Germany}

\author{A.~V.~Chumak}
\affiliation{Fachbereich Physik and Forschungszentrum OPTIMAS, Technische Universit\"at Kaiserslautern, 67663
Kaiserslautern, Germany}

\author{V.~I.~Vasyuchka}
\affiliation{Fachbereich Physik and Forschungszentrum OPTIMAS, Technische Universit\"at Kaiserslautern, 67663
Kaiserslautern, Germany}

\author{A.~A.~Serga}
\affiliation{Fachbereich Physik and Forschungszentrum OPTIMAS, Technische Universit\"at Kaiserslautern, 67663
Kaiserslautern, Germany}

\author{B.~Obry}
\affiliation{Fachbereich Physik and Forschungszentrum OPTIMAS, Technische Universit\"at Kaiserslautern, 67663
Kaiserslautern, Germany}

\author{H.~Schultheiss}
\altaffiliation{Current adress: Materials Science Division and Center for Nanoscale Materials, Argonne National Laboratory, Argonne, Illinois 60439, USA}
\affiliation{Fachbereich Physik and Forschungszentrum OPTIMAS, Technische Universit\"at Kaiserslautern, 67663
Kaiserslautern, Germany}

\author{P.~A.~Beck}
\affiliation{Fachbereich Physik and Forschungszentrum OPTIMAS, Technische Universit\"at Kaiserslautern, 67663
Kaiserslautern, Germany}

\author{A.~D.~Karenowska}
\affiliation{Department of Physics, Clarendon Laboratory, University of Oxford, OX1 3PU Oxford, United Kingdom}

\author{E.~Saitoh}
\affiliation{Institute for Materials Research, Tohoku University, Sendai 980-8577, Japan}

\author{B.~Hillebrands}
\affiliation{Fachbereich Physik and Forschungszentrum OPTIMAS, Technische Universit\"at
Kaiserslautern, 67663 Kaiserslautern, Germany}

\date{\today}

\begin{abstract}

It is demonstrated that upon pulsed microwave excitation, the temporal behavior of a spin-wave induced inverse spin Hall voltage in a magnetic insulator-nonmagnetic metal structure is distinctly different from the temporal evolution of the directly excited spin-wave mode from which it originates. The difference in temporal behavior is attributed to the excitation of long-lived secondary spin-wave modes localized at the insulator-metal interface.

%The time dependence of an inverse spin Hall  (iSHE) voltage in a magnetic insulator (yttrium iron garnet,
%YIG)-nonmagnetic metal (platinum, Pt) structure is studied. A flow of spin-polarized electrons - the spin-dependent
%scattering of which originates the iSHE voltage - is pumped into the platinum layer by microwave frequency spin waves
%excited in the YIG. The temporal behavior of the iSHE voltage and the spin-wave intensity are compared. It is revealed
%that the iSHE voltage evolves slowly in comparison with the externally driven spin-wave mode. We suggest that this is
%explained by the fact that secondary spin waves created by two-magnon  scattering of the externally excited spin-wave
%mode make a strong contribution to the iSHE signal.
\end{abstract}

\maketitle

Over the last decade, the field of spintronics has risen to some prominence. Spintronics is concerned with the development of devices which exceed the performance and energy efficiency of conventional charge-based electronics by exploiting the electron's spin degree of freedom \cite{Kajiwara2010,Awschalom,Zutic,Gregg}. The information currency in spintronic systems is spin angular momentum. Traditional spintronic architectures rely on the electron based spin-transport, however, spin angular momentum can also be transferred by magnons, the quanta of spin waves (collective excitations of the spin lattice of a magnetic material). Magnon physics opens doors to insulator-based spintronic devices which operate with pure {\it spin} currents entirely decoupled from {\it charge} carriers \cite{Kajiwara2010,Uchida2010}. Spin waves in magnetic insulators can propagate over macroscopic distances many orders of magnitude longer than the spin diffusion lengths typical of metallic and semiconductor materials \cite{YIG_magnonics,Realization_SW_logic,Wang_logic}. Spin pumping (which transforms spin waves into spin polarized electron currents) and the inverse spin Hall effect (iSHE) (which converts spin polarized electron currents into conventional charge currents) are two physical mechanisms of fundamental importance to the emerging field of `magnon spintronics'. The temporal characteristics of these phenomena will ultimately determine the operational speeds of magnon spintronic devices \cite{Stern}.

\begin{figure}[t]
\includegraphics[width=0.75\columnwidth]{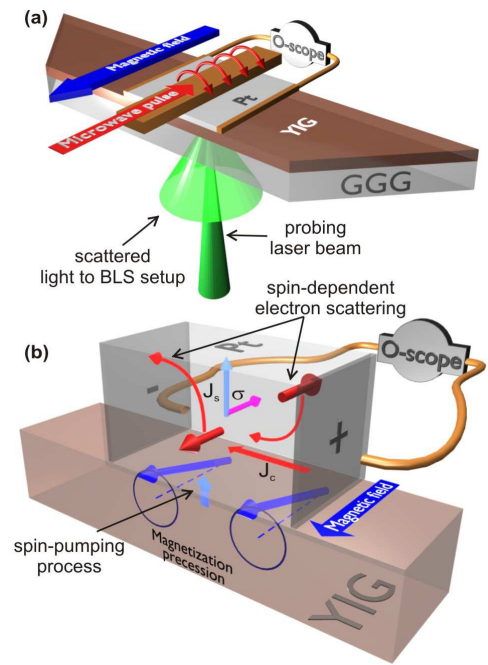}
\caption{\label{fig:aufbau} (Color online) (a) Schematic illustration of the experimental setup.
(b) Spin pumping scheme and resulting inverse spin Hall effect.}
\end{figure}

In this Letter we report our findings relating to the temporal behavior of an externally excited spin-wave pulse and a resulting inverse spin Hall voltage in a magnetic insulator-nonmagnetic metal bi-layer. We demonstrate that there are pronounced differences in the temporal evolution of the two signals and show that these differences may be attributed to the excitation of {\it secondary} short-wavelength spin-waves localized at the insulator-metal interface via two-magnon scattering of the externally excited mode.

%In this letter we report on our findings of the temporal behavior of an inverse spin Hall voltage in a magnetic insulator -- nonmagnetic metal bi-layer and compare it with the evolution of the spin-wave spectrum from which it originates. We show that the iSHE voltage only partially mirrors the intensity of the externally driven spin-wave mode: the iSHE pulsed signal has a tail due to the pumping effect of {\it secondary} noncoherent spin waves created by two-magnon scattering of the initially %excited mode.

A platinum (Pt) coated  yttrium iron garnet (YIG) film was used in our experiments. YIG single crystal films have the
smallest known spin-wave damping \cite{YIG_magnonics,Gurevich}. As a result, magnon currents can be observed in
YIG over centimeter distances \cite{Kajiwara2010}. Electron scatter in Platinum is strongly spin-dependent making it an attractive material for iSHE voltage generation \cite{Hirsch,Bauer2}. The experimental setup is illustrated schematically
in Fig.~\ref{fig:aufbau}(a). It comprises a 2.1~{$\mu$m} thick YIG stripe with a 10~nm thick $3 \times 3$~mm$^{2}$ Pt
layer deposited on the top. The edge-to-edge resistance of the Pt square is 54~Ohms. To ensure good impedance matching and thus
minimal distortion of detected signals, the Pt square was connected to the 50~Ohm input of a voltage measuring
instrument (see Fig.~\ref{fig:aufbau}(a)) by a 50~Ohm coaxial cable.

Magnetization precession was excited in the YIG layer by a microwave current applied to a copper microstrip line of width $600$~$\mu$m above the Pt layer. The line was isolated from the Pt by a thin dielectric coating of cyan-acrylate. Two excitation geometries were investigated. In the first, the microstrip line was placed across the YIG stripe (Fig.~\ref{fig:aufbau}(a)), in the second (not shown) it was orientated parallel to it. The temporal behavior of the iSHE voltages observed in the two cases was near identical. Accordingly, we present here only data collected using the first geometry.

\begin{figure}[t]
\includegraphics[width=0.95\columnwidth]{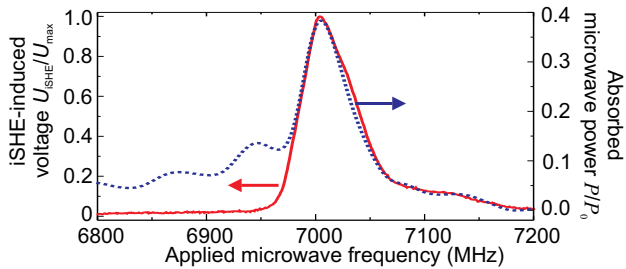}
\caption{\label{fig:Spec} (Color online)
The inverse spin Hall voltage (solid red line) and the absorbed microwave power (blue dashed line) are shown as functions of the applied microwave frequency. Data corresponds to a bias magnetic field of $H_{0} = 1820$~Oe. The maximum iSHE voltage $U_\textrm{max} = 60$~$\mu$V is observed at the ferromagnetic resonance frequency $f_{0} = 7$~GHz. A decrease in the spin-pumping efficiency for modes excited at frequencies smaller than $f_{0}$ is evident (see Ref.~\cite{Sandweg}).}
\end{figure}

The spin-wave modes excited in the YIG film were detected via time-resolved Brillouin light scattering (BLS) spectroscopy \cite{Demokritov2001}. The BLS laser probe beam was focused on the YIG/Pt sample and the intensity of inelastically scattered light (which is directly proportional to the intensity of the scattering spin wave) was analyzed on a 500~ps timescale. For the time-resolved voltage measurements we used a wideband (DC to 200~MHz) voltage amplifier FEMTO DHPVA-200 and a 300~MHz bandwidth Agilent DSO6034A oscilloscope.

Our experiments were performed in the following fashion: a magnetizing field $\textbf{H}_{0}$ was applied across the YIG stripe and magnetization precession was driven by a magnetic field $\textbf{h}(t)$ induced by a microwave current applied to the microstrip line. Under these conditions, due to the spin pumping effect of the precessing magnetization at the YIG/Pt interface, a spin polarized electron current $\textbf{J}_{\mathrm{s}}$ is produced in the Pt layer \cite{Kajiwara2010,Bauer2,Mizukami}. As a consequence of spin-dependent electron scattering in the Pt layer \cite{Hirsch} this spin polarized current leads in turn to a conventional charge current $\textbf{J}_{\mathrm{c}}$ and thus a charge accumulation transverse both to $\textbf{H}_{0}$ and $\textbf{J}_{\mathrm{s}}$. Accordingly, an iSHE voltage $U_\textrm{iSHE}$ appears across the Pt square (see Fig.~\ref{fig:aufbau}~(b)). As $\textbf{J}_{\mathrm{c}} \propto \textbf{J}_{\mathrm{s}} \times \boldsymbol{\sigma}$ the polarity of the iSHE voltage can be changed by changing the polarization of the spin current $\boldsymbol{\sigma}$ via the static magnetization of the YIG stripe.

In order to increase the dynamic range of the time-resolved iSHE and BLS measurements we supplied a moderately high microwave power to the microstrip line ($P_{0} = 100$~mW). So as to avoid possible caloric effects, driving microwave pulses of 1~$\mu$s duration were applied with a repetition rate of 10~$\mu$s. The rise and fall times of these pulses were less than 5~ns.

First of all, in order to confirm the origin of the voltage $U_\textrm{iSHE}$ we observed, we verified that its polarity was indeed dependent on the polarity of the magnetization direction. To additionally corroborate our results and to rule out parasitic effects (for example electromagnetic induction in the Pt stripe) we also tested a structure with a nonmagnetic insulator (gadolinium gallium garnet, GGG) in the place of the YIG. No voltage was detected. Thus, we are able to say with confidence that the voltages we observed were due to the iSHE.

The absorbed microwave power (directly proportional to the intensity of the excited spin waves) and the iSHE voltage were measured as the applied microwave frequency was varied for a magnetizing field $H_{0} = 1820$~Oe (see Fig.~\ref{fig:Spec}). The maximum microwave absorption was recorded for the spin-wave mode excited at the frequency of ferromagnetic resonance (FMR) $f_0=7$~GHz. The observed FMR linewidth was around 50~MHz (corresponding to $2\Delta H_{\mathrm{FMR}} = 17$~Oe). This value is significantly larger than the FMR linewidth of $0.6$~Oe quoted by the producers of the YIG film. The apparent discrepancy is due to the fact that in our
experiments the FMR mode is strongly coupled to the exciting microstrip line, and thus the dominant dissipative
mechanism is not the small magnetic damping but the large radiation loss back into the line.

The 50~MHz  bandwidth of the microstrip-loaded FMR was wide enough to contain the frequency spectrum of the driving
microwave pulse of 1~$\mu$s duration without significant distortion. In order to determine the distortion level the
convolution of the Fourier transformed input pulse and the spectrum of the absorbed microwave power was calculated.
Using the inverse Fourier transformation of this convolution, we obtained the calculated `tailing' of the spin-wave
pulse edges, which was smaller than 20~ns. As the measured iSHE voltage originates directly from the spin-wave
amplitude at the YIG/Pt interface (Fig.~\ref{fig:Spec}) a similar tail was expected to be observed on the $U_\textrm{iSHE}$ pulse. However, the real temporal behavior both of the spin-wave amplitude and iSHE voltage proved to be much more complex. Note that
time-resolved BLS measurements performed at $f_0=7$~GHz established that excitation delays between different sample
points were always smaller than $1$~ns, confirming the excitation of quasi-uniform FMR rather than spin-wave modes
traveling away from the microstrip line. Therefore, no time delays associated with a nonzero spin-wave propagation
time through the Pt area appear in the iSHE signal.

\begin{figure}[b]
\includegraphics[width=0.95\columnwidth]{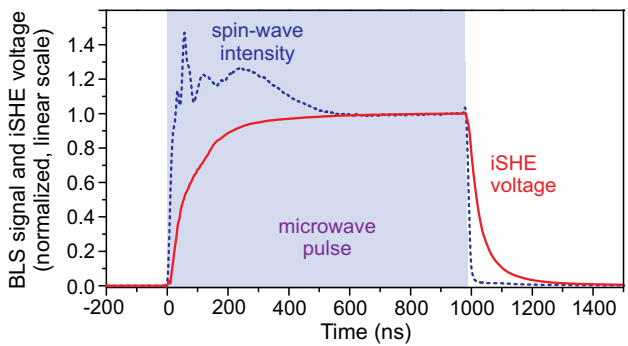}
\caption{\label{fig:comp} (Color online) Comparison of the normalized spin-wave signal measured with Brillouin light
scattering spectroscopy (blue curve) and the iSHE voltage (red curve). The maximum iSHE voltage $U_\textrm{max}$ is
60~$\mu$V.}
\end{figure}

The time resolved measurements were performed at $f_{0} = 7$~GHz where the iSHE voltage was maximum. In
Fig.~\ref{fig:comp} the time profiles of the spin-wave and voltage pulses are compared. The spin-wave intensity (blue
dashed line in Fig.~\ref{fig:comp}) increases rapidly when the microwave pulse is applied at $t = 0$~ns. After some
oscillations corresponding to a nonlinear transition process (common for relatively high spin-wave intensities
\cite{Lvov}), an equilibrium value is reached. When the microwave pulse is switched off at $t = 1000$~ns, the BLS
signal decreases rapidly. The measured iSHE voltage is shown by a red solid line in Fig.~\ref{fig:comp}. It is clear
that the rise and fall times of the iSHE voltage are appreciably longer than those corresponding to the spin-wave
intensity.

\begin{figure} [t]
\includegraphics[width=0.95\columnwidth]{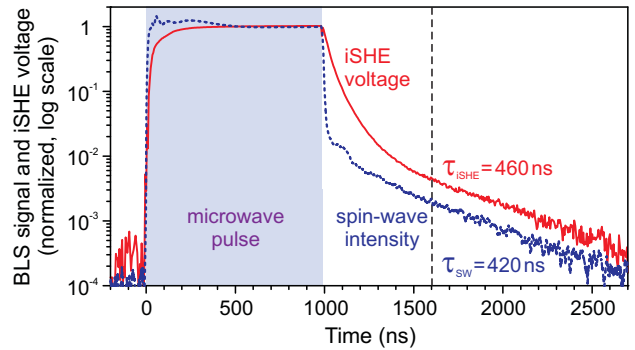}
\caption{\label{fig:comp2} (Color online)
Spin-wave intensity and iSHE voltage as a function of time (logarithmic scale).}
\end{figure}

In order to unpick the peculiarities of the temporal evolution of the iSHE signal and to relate these to the dynamics
of the precessing magnetization, we show the waveforms plotted on logarithmic scale in Fig.~\ref{fig:comp2}. The data
of Fig.~\ref{fig:comp2} has several features which warrant clarification:

Firstly, the falling slopes of both the iSHE voltage and of the spin-wave intensity  are nonexponential (note the
logarithmic scale).

Secondly, the measured iSHE voltage rises and decreases much more slowly than  the spin-wave intensity; for example,
during the first 50~ns after the driving microwave pulse has been switched off, the iSHE voltage decreases only by factor of
3 whereas the spin-wave intensity falls by factor of 50.

Thirdly, for $t > 1600$~ns the iSHE voltage and spin-wave intensity decay  exponentially and their fall times
$\tau_{\mathrm{iSHE}}^{}$ and $\tau_{\mathrm{SW}}^{}$ are very similar; 460~ns and 420~ns respectively.

These facts can be understood if one assumes that---rather than solely the externally  driven spin-wave mode---many
modes contribute to the iSHE voltage. In order to illustrate how such a model fits with the experimental data we
consider for simplicity only two groups of modes. The first group corresponds to the FMR directly
excited by the microstrip line. This group is characterized by a large amplitude (determined by the applied microwave signal)
and a high decay rate due to its strong coupling to the microstrip antenna (see Fig.~\ref{fig:Spec}). A second mode group is excited {\it indirectly} via two-magnon scattering of the first group by defects and inhomogeneities in the YIG film. This well known mechanism results in the redistribution of energy from the externally excited uniform mode into dipolar-exchange spin-wave (DESW) modes with wavelengths determined by sizes of the scatterers (typically 1~$\mu$m in high quality YIG samples) \cite{YIG_magnonics,Gurevich,Sparks,Melkov1,Melkov2}. Due to their extremely short wavelengths, the DESW modes are entirely decoupled from the microstrip line and their relaxation ($2\Delta H_{\mathrm{DESW}} = 0.1-0.2$~Oe \cite{Gurevich,Melkov1,Melkov2}) is dominated by the intrinsic magnetic damping of the film \cite{Gurevich}. Since the impurities and inhomogeneities are concentrated close to the YIG film surfaces, these modes are localized near the YIG/Pt interface.

The weak DESW group cannot be resolved by BLS on the background of the strong uniform  FMR mode (in addition, the BLS setup is less sensitive to short-wavelength spin waves) and thus is visible only after the microwave pulse has been switched off and the fast-relaxing FMR precession has decayed. It follows that the spin-wave fall time
$\tau_{\mathrm{SW}}^{}$ of 420~ns measured for  $t > 1600$~ns is determined {\it only} by the relaxation of the DESW
modes. Therefore, for $t > 1600$~ns, the long-lived DESW modes are the sole contributors to the BLS signal and the iSHE
voltage. As a result, the slopes $\tau_{\mathrm{iSHE}}^{}$ and $\tau_{\mathrm{SW}}^{}$ are approximately the same. The
value of the fall time in this region of the signal is a direct evidence for the dipolar-exchange nature of these
waves: it corresponds to a resonance curve linewidth of $2\Delta H_{\mathrm{DESW}} = 0.14$~Oe which fits very well with
literature data on DESW relaxation.

We suggest that the DESW modes, in spite of their small amplitude, make a relatively more significant contribution
to the iSHE voltage than the directly excited FMR mode due to their localization close to the YIG/Pt interface. As a
result, after the excited spin-wave intensity maximum, the iSHE voltage continues to grow (for times $t < 500$~ns). The
same effect is visible after the microwave pulse has been switched off: the iSHE voltage continues to be generated by the
long-lived DESW group.

The model we propose accounts well for the main features of the experimental results even in the case that we consider contributions from only two spin-wave groups to the iSHE voltage signal (the slowest and the
fastest relaxing). For quantitative analysis of the transition regions (times $t < 500$~ns and  $1000$~ns $ <
t < 1600$~ns) the contributions of spin-wave modes having intermediate life times must be taken into account (for
example, in order to fit the entire iSHE slope for time interval $t > 1000$~ns three exponential functions are
required). Experimentally, these modes can be excited both as a result of elastic two-magnon scattering and due to
nonlinear four-magnon scattering of the FMR mode. Additional experiments performed at different microwave powers (in
the range from 25 mW to 1 W) showed slight variations of the iSHE slopes for times $t < 500$~ns and  $1000$~ns $ < t <
1600$~ns which can be attributed to the effect of nonlinearity. Nevertheless, we emphasize that the qualitative
behavior of the voltage pulse as well as its fall time for $t > 1600$~ns remains the same regardless of the FMR mode
power.

In summary, we report that the iSHE voltage increases and decreases significantly more slowly than the intensity of the pulsed externally driven uniform spin-wave mode from which it originates. From the complex temporal behavior of both signals we have observed we conclude that {\it indirectly} excited short-wavelength dipolar-exchange spin waves participate in spin pumping at the metal-insulator interface as well as {\it directly} excited uniform precession. The surface localization of the scattered dipolar-exchange modes means that despite their low amplitude and incoherent character, they make a
substantial contribution to the iSHE voltage signal. In addition, we can conclude that iSHE voltage signal delays in magnetic insulator-nonmagnetic metal structures are dominated by spin-wave dynamics in the insulator, rather than electron dynamics in the metal.

%Our studies show that transient processes in magnetic insulator - non-magnetic metal structures are mainly determined
%by delays in the spin-wave system and much longer than spin diffusion time or any other experimentally relevant time
%constants.
%
%The equality of the slowest times in iSHE voltage with the spin-wave intensity give us a possibility to conclude that
%the slowest processes which determine the delays of iSHE voltage in magnetic insulator - non-magnetic metal structures
%have place inside a spin-wave system in YIG rather than spin accumulation or spin diffusion times in Pt.

We thank G.~E.~W. Bauer for the valuable discussions and the Nano+BioCenter, TU Kaiserslautern, for technical support. A.~D.~K. is grateful for the support of Magdalen College, Oxford. B.~O. would like to acknowledge the DFG for support within the Graduiertenkolleg~792.


\begin{thebibliography}{19}
%
\bibitem{Kajiwara2010} Y. Kajiwara, K. Harii, S. Takahashi, J. Ohe, K. Uchida, M. Mizuguchi, H. Umezawa, H. Kawai, K. Ando, K. Takanashi, S. Maekawa, and E. Saitoh,
    %Transmission of electrical signals by spin-wave interconversion in a magnetic insulator,
    Nature \textbf{464}, 262 (2010).

\bibitem{Awschalom} S. A. Wolf, D. D. Awschalom, R. A. Buhrman, J. M. Daughton, S. von Moln\'{a}r, M. L. Roukes, A. Y. Chtchelkanova, and D. M. Treger,
        %Spintronics: a spin-based electronics vision for the future,
        Science \textbf{294}, 5546 (2001).

\bibitem{Zutic} I. \v{Z}utic, J. Fabian, and S. Das Sarma,
        %Spintronics: Fundamentals and applications,
        Rev. Mod. Phys. \textbf{76}, 2 (2004).

\bibitem{Gregg} J. F. Gregg,
        %Spintronics: A growing science,
        Nature Mater. \textbf{6}, 798 (2007).

%\bibitem{Bader} S. D. Bader and S. S. P. Parkin,
        %Spintronics,
        %Annual Review of Condensed Matter Physics \textbf{1}, 71 (2010).

\bibitem{Uchida2010} K. Uchida, J. Xiao, H. Adachi, J. Ohe, S. Takahashi, J. Ieda, T. Ota, Y. Kajiwara, H. Umezawa, H. Kawai, G. E. W. Bauer, S. Maekawa, and E. Saitoh,
        %Spin Seebeck insulator,
        Nature Mater. \textbf{9}, 894 (2010)

\bibitem{YIG_magnonics} A. A. Serga, A. V. Chumak, and B. Hillebrands,
        %YIG magnonics,
        J. Phys. D: Appl. Phys. \textbf{43}, 264002 (2010).

%\bibitem{YIG_saga} V. Cherepanov, I. Kolokolov, and V. L'vov,
        %The saga of YIG: spectra, thermodynamics, interaction and relaxation of magnons in a complex magnet,
        %Phys. Rep.---Rev. Sec. Phys. Lett. \textbf{229}, 81 (1993).

\bibitem{Realization_SW_logic} T. Schneider, A. A. Serga, B. Leven, B. Hillebrands, R. L. Stamps, and M. P. Kostylev,
        %Realization of spin-wave logic gates,
        Appl. Phys. Lett. \textbf{92}, 022505 (2008).

\bibitem{Wang_logic} A. Khitun, M. Bao, J. Lee, K. Wang, D. W. Lee, and S. Wang,
        %Spin Wave Based Logic Circuits,
        Materials Research \textbf{998} (2007).

\bibitem{Stern} N.P. Stern, D. W. Steuerman, S. Mack, A.C Gossard, and D.D. Awschalom
        %Time-resolved dynamics of the spin Hall effect
        Nature Physics \textbf{4}, 843 (2008).

\bibitem{Gurevich} A. G. Gurevich and G. A. Melkov,
        \textit{Magnetization Oscillations and Waves} (CRC, New York, 1996).

\bibitem{Sandweg} C. W. Sandweg, Y. Kajiwara, K. Ando, E. Saitoh, and B. Hillebrands
        %Enhancement of the spin pumping efficiency by spin wave mode selection
        Appl. Phys. Lett. \textbf{97}, 252504 (2010).

\bibitem{Hirsch} J. E. Hirsch,
        %Spin Hall effect,
        Phys. Rev. Lett. \textbf{83}, 1834 (1999).

%\bibitem{Zhang} S. Zhang,
        %Spin hall effect in the presence of spin diffusion,
        %Phys. Rev. Lett. \textbf{85}, 393 (2000).

%\bibitem{Bauer} A. Brataas, Y. Tserkovnyak, G. E. W. Bauer, and B. I. Halperin,
        %Spin battery operated by ferromagnetic resonance,
        %Phys. Rev. B \textbf{66}, 060404(R) (2002).

\bibitem{Bauer2} Y. Tserkovnyak, A. Brataas, and G. E. W. Bauer,
        %Spin pumping and magnetization dynamics in metallic multilayers,
        Phys. Rev. B \textbf{66}, 224403 (2002).

\bibitem{Demokritov2001} S. O. Demokritov, B. Hillebrands, and A. N. Slavin,
        %Brillouin light scattering studies of confined spin waves: linear and nonlinear confinement,
        Phys. Rep. \textbf{348}, 441 (2001).

%\bibitem{BLS} O. B\"uttner, M. Bauer, S. O. Demokritov, B. Hillebrands, Y. S. Kivshar, V. Grimalsky, Yu. Rapoport, and A. N. Slavin,
        %Linear and nonlinear diffraction of dipolar spin waves in yttrium iron garnet films observed by space- and time-resolved Brillouin light scattering,
        %Phys. Rev. B \textbf{61}, 11576 (2000).

%\bibitem{Silsbee} R. H. Silsbee, A. Janossy, and P. Monod,
        %Coupling between ferromagnetic and conduction-spin-resonance modes at a ferromagnetic - normal-metal interface,
        %Phys. Rev. B \textbf{19}, 4382 (1979).

\bibitem{Mizukami} S. Mizukami, Y. Ando, and T. Miyazaki,
        %Effect of spin diffusion on Gilbert damping for a very thin permalloy layer in Cu/permalloy/Cu/Pt films,
        Phys. Rev. B \textbf{66}, 104413 (2002).

%\bibitem{Dyakonov} M. I. Dyakonov and V. I. Perel,
        %Current-induced spin orientation of electrons in semiconductors,
        %Phys. Lett. A \textbf{35}, 6 (1979).

%\bibitem{GangulyWebb} A. K. Ganguly and D. C. Webb,
        %Microstrip Excitation of Magnetostatic Surface Waves: Theory and Experiment
        %IEEE Trans. Microwave Theor. and Techn. \textbf{MTT-23}, 998 (1975).

%\bibitem{Stancil} D. D. Stancil and A. Prabhakar,
        %\textit{Spin waves - Theory and Applications} (Springer, 2009).

\bibitem{Lvov} V. S. L'vov, \textit{Wave Turbulence under Parametric Excitations: Applications to Magnetics} (Springer, Berlin, 1994).

%\bibitem{Cottam} M. G. Cottam and D. J. Lockwood,
        %\textit{Light scattering in magnetic solids} (Wiley-Interscience, 1986).

\bibitem{Sparks} M. Sparks,
        \textit{Ferromagnetic Relaxation Theory} (McGraw-Hill, New York, 1964).

\bibitem{Melkov1} G. A. Melkov, V. I. Vasyuchka, Yu. V. Kobljanskyj, and A. N. Slavin,
        %Wave-Front reversal in a medium with inhomogeneities and an anisotropic wave spectrum,
        Phys. Rev. B \textbf{70}, 224407 (2004).

\bibitem{Melkov2} G. A. Melkov, A. D. Dzyapko, A. V. Chumak, and A. N. Slavin,
        %Two-magnon relaxation reversal in ferrite spheres,
        J. Exp. and Theor. Phys. \textbf{99}, 1193 (2004).

%\bibitem{DESW} Y. V. Kobljanskyj, G. A. Melkov, V. S. Tiberkevich, V. I. Vasyuchka, and A. N. Slavin,
        %Microwave signal processing using dipole-exchange spin waves
        %J. Appl. Phys. \textbf{93}, 8594-8597 (2003).

%\bibitem{Melkov3} G. A. Melkov, V. I. Vasyuchka, A. V. Chumak, and A. N. Slavin,
        %Double-wave-front reversal of dipole-exchange spin waves in yttrium-iron garnet films
        %J. Appl. Phys. \textbf{98}, 074908 (2005).

%\bibitem{Bass} J. Bass and W. P. Pratt Jr.
%                %Spin-diffusion lengths in metals and alloys, and spin-flipping at metal/metal interfaces: an experimentalist's critical review
%                J. Phys.: Condensed Matter \textbf{19}, 183201 (2007).
%
%\bibitem{Maekawa} S. Takahashi and S. Maekawa
%                %Spin acurrent, spin accumulation and spin Hall effect
%                Sci. Technol. Adv. Mater. \textbf{9}, 014105 (2008)
%
%

\end{thebibliography}
\end{document}